\documentclass{pasa}%

\title[Automatic Recognition of Sunspots in HSOS Images]{Automatic Recognition of Sunspots in HSOS Full-Disk Solar Images}
\author[Cui Zhao et al.]{Cui Zhao$^1$, GangHua Lin$^1$, YuanYong Deng$^1$ \and Xiao Yang$^1$\\
\affil{$^1$Key Laboratory of Solar Activity, National Astronomical Observatories, Chinese Academy of Sciences, 20A Datun Road, Beijing 100012, China}}%

\jid{PASA}
\jyear{\the\year}

\usepackage[authoryear]{natbib}
\bibpunct{(}{)}{;}{a}{}{,}
\usepackage{aas_macros}
\usepackage{hyperref}
\hypersetup{colorlinks,citecolor=blue,linkcolor=blue,urlcolor=blue}

\begin{document}

\begin{abstract}
A procedure is introduced to recognise sunspots automatically in solar full-disk photosphere images obtained from Huairou Solar Observing Station, National Astronomical Observatories of China. The images are first pre-processed through Gaussian algorithm. Sunspots are then recognised by the morphological Bot-hat operation and Otsu threshold. Wrong selection of sunspots is eliminated by a criterion of sunspot properties. Besides, in order to calculate the sunspots areas and the solar centre, the solar limb is extracted by a procedure using morphological closing and erosion operations and setting an adaptive threshold. Results of sunspot recognition reveal that the number of the sunspots detected by our procedure has a quite good agreement with the manual method. The sunspot recognition rate is 95\% and error rate is 1.2\%. The sunspot areas calculated by our method have high correlation (95\%) with the area data from USAF/NOAA.
\end{abstract}
\begin{keywords}
methods: data analysis -- Sun: activity --  sunspots -- techniques: image processing
\end{keywords}
\maketitle%
\section{INTRODUCTION}
\label{sec:intro}

Sunspots are the most obvious phenomenon on the solar surface with high magnetic fields. Researches have shown they are evidently related to the solar cycle and other solar activities (flare, filament, CME etc.) \citep{WangHM13}. Sunspot properties (including size, location, number, magnetic classification etc.) are usually applied to predict solar activities and monitor the space environment \citep{DuZL12}. Therefore, acquiring their properties is significative. Accurate recognition is essential for extracting sunspots and then able to acquire their properties. In early stage, it was manually recognised \citep{Steinegger96}, which is inefficient and not real-time, could not deal with the statistically requirements for numbers of data and needs to monitor space environment in real-time. Automatic recognition with high precision is therefore expected.

Several methods have been proposed to recognise sunspots. \citet{Zharkov05} used edge-detection method and a local threshold to find the sunspot candidates, and a median filter was employed to remedy initial over segmentation of images. \citet{Curto08} applied erosion and eroded gradient transformation to the detection of the solar limb, and then top hat operator was used to obtain valley regions on the solar disk. \citet{Watson09} employed the morphological operations and then an intensity threshold to obtain candidates for the sunspot regions. \citet{Djafer12} adopted wavelet analysis to detect sunspots and the solar limb on Ca\,{\small II} K1 Meudon images. \citet{Goel14} adopted a method called level set to detect and track sunspots.

Huairou Solar Observing Station (HSOS) is one of the key stations of the National Astronomical Observatories, Chinese Academy of Sciences. It has been observing sunspots for many years \citep{ZhangHQ07}. One of its telescopes can acquire the full-disk photospheric data and vector magnetic fields at the same time. Recognising sunspots based on these data will provide sunspot geometrical and magnetic properties, and is meaningful for sunspot study. Thus, a procedure for automatic recognition of sunspots in HSOS full-disk solar images is expected.

A large number of above methods had been tested to do with that, but did not work well in HSOS image. This is mainly due to the following reasons: (1) affected by ground atmosphere and with the limited diameter of telescope, some sunspots umbra and penumbra in the HSOS images are inseparable; (2) in the year of 2011, as the problem of the telescope, these are instrument noises in HSOS images. Above methods do not consider the interference by them. So in order to extract sunspots accurately from HSOS full-disk images, we propose a new automatic recognition procedure, by which a catalogue of sunspot properties is expected to generate.

This paper is arranged as follows: in Section 2, we briefly introduce the data that are used. Related theories and tools are discussed in Section 3. In Section 4, we give a detailed description of the procedure to identify the sunspots. We verify our procedure and make a discussion in Section 5. The summary is in Section 6.

\section{DATA}

HSOS has been equipped a 10-cm full-disk vector magnetograph since 2006 that is able to measure full-disk vector magnetic fields at Fe\,{\small I} 5324.19 \AA\ and obtain photospheric images simultaneously. The image frame size is 992$\times$1004, with the pixel resolution of $2'\times 2'$ and spatial resolution better than 5 arcsec. Sunspots can be observed in the images. Combined with the synchronous vector magnetic field data, automatic identification of sunspots can get their geometrical and magnetic properties.

\section{RELATED TOOLS FOR SUNSPOT RECOGNITION}

Before introducing sunspots recognition procedure, the related tools will be introduced first.

\subsection{Mathematical morphology}

Mathematical morphology is a tool for extracting image components that are useful for representation and description \citep{Haralick87}. The basic idea is by means of the structure of a certain morphology to measure and extract the corresponding shape of an image. Compared with the differential operator to extract edges, this method has following advantages: it is not so sensitive to noise as differential operator, meanwhile the recognised edges are smooth, continuous, and have less breakpoints.

The two basic morphological set transformations are erosion and dilation. These transformations involve the interaction between an image I (the object of interest) and a structuring set B, called the SE. A lot of other morphological operations are derived from them. In the following, erosion, dilation, closing, opening, and Bot-hat transformation will be introduced.

(1) Erosion

To obtain the minimum grey value of the original image I minus the one of the block B, erosion operation is defined as follows:
\[
I\ominus B = \min\{I(x+x',y+y') - B(x',y')|(x',y')\in D_b\}.
\]
Here $D_b$ is the neighborhood block.

(2) Dilation

Dilation is to obtain the maximum grey value of the original image plus the one of block. It is defined as follows:
\[
I\oplus B = \max\{I(x-x',y-y') + B(x',y')|(x',y')\in D_b\}.
\]
Here $D_b$ mean the same as in (1).

(3) Closing

Closing operation is a process of dilation followed by erosion. It is defined as follows:
\[
I\bullet B = (I\oplus B)\ominus B.
\]
It is employed to fill small holes, with objects connect the adjacent objects and smooth the boundaries of the images.

(4) Opening

Opening generally smooths a contour in an image, breaking narrow isthmuses and eliminating thin protrusions. It is defined as follows:
\[
I\circ B = (I\ominus B)\oplus B.
\]

(5) Bot-hat transformation

Bot-hat transformation is a subtraction of the original image and the closing image, it is defined as follows:
\[
T = I - (I\oplus B)\ominus B
\]
$T$ shows a grey valley in the original image, and highlights the boundaries between connected objects. It is able to extract dark pixels from a bright background.

\subsection{Otsu algorithm}

Otsu algorithm is a way to search an adaptive threshold, proposed by \citet{Otsu79}, whose idea is to divide an image into two parts: background and objective with a sharp discrepancy. The more different the two parts are, the sharper the discrepancy is. Part of the objectives wrongly divided into the background will make the discrepancy smaller. Therefore, the sharpest discrepancy means minimum probability of erroneous division. The calculation principle is as follows.

Now suppose that we dichotomise the pixels of the image $I(x,y)$ into two classes (objects and background) by a threshold $T$, the objects and background pixels numbers are $N_1$ and $N_2$, respectively. The image size is indicated by $M\times N$, the percentage of the objects pixel number is $\omega_1$, the percentage of the background ones is $\omega_2$, then we can easily verify the following relation:
\begin{align*}
& \omega_1 = \frac{N_1}{M\times N}  \tag*{(1)}  \\
& \omega_2 = \frac{N_2}{M\times N}  \tag*{(2)}  \\
& N_1 + N_2 = M\times N  \tag*{(3)}  \\
& \omega_1 + \omega_2 = 1  \tag*{(4)}  \\
\end{align*}
$\omega_1$ and $\omega_2$ are the objects and background area probabilities, respectively \citep{sezgin2004survey}.

The objects and the background average grey values are shown by $\mu_1$ and $\mu_2$, respectively. The image average grey value is then by $\mu$
\[
\mu = \mu_1 \times \omega_1 + \mu_2 \times \omega_2.  \tag*{(5)}
\]
It is the total mean level of the original picture.

$g$ is given by
\[
g = \omega_1\times (\mu - \mu_1)^2 + \omega_2\times(\mu - \mu_2)^2.  \tag*{(6)}
\]
It represents the variances between objects and background.

Substituting Equation (5) into (6) gives
\[
g = \omega_1 \times \omega_2 \times (\mu_1 - \mu_2)^2.  \tag*{(7)}
\]
Iterating all possible pixel values at the threshold $T$, the value which makes $g$ the biggest will be the result.

\section{RECOGNITION PROCEDURE}

After the introduction to the above tools, sunspot recognition procedure based on them could be described.

In order to recognise sunspots and calculate their parameters (sunspot areas, position and so on), two steps are carried out. In the first step, the solar limb is extracted from the solar disk and the solar centre and radius are fixed. In the second, to recognise sunspots, morphological Bot-hat operation and local threshold are employed. Over segmentation of sunspots is eliminated by limitation in sunspot properties.More details are introduced in the following sections.

\subsection{Extraction of solar limb}

\begin{figure*}
\begin{center}
\includegraphics[width=12pc]{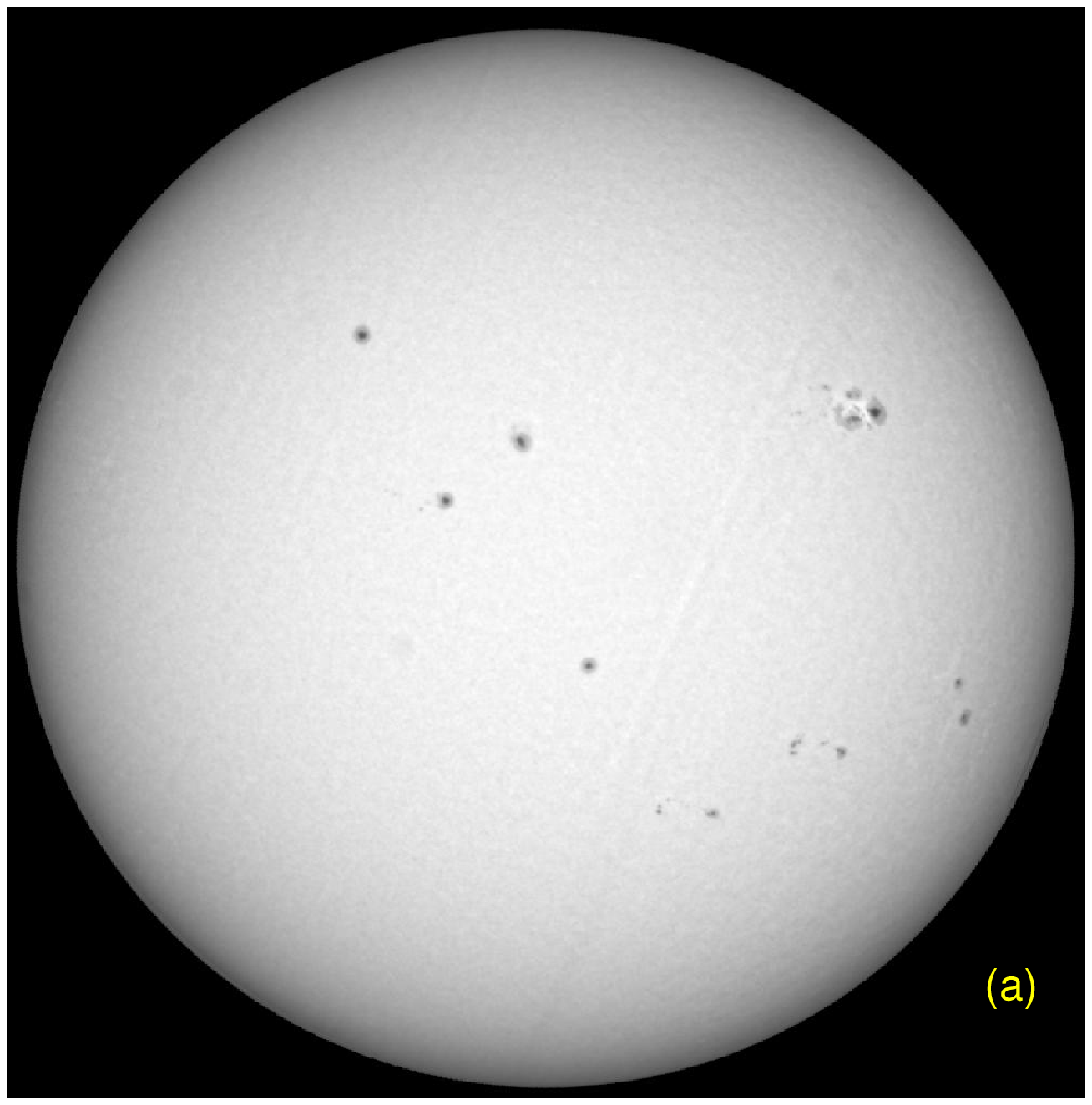}
\includegraphics[width=12pc]{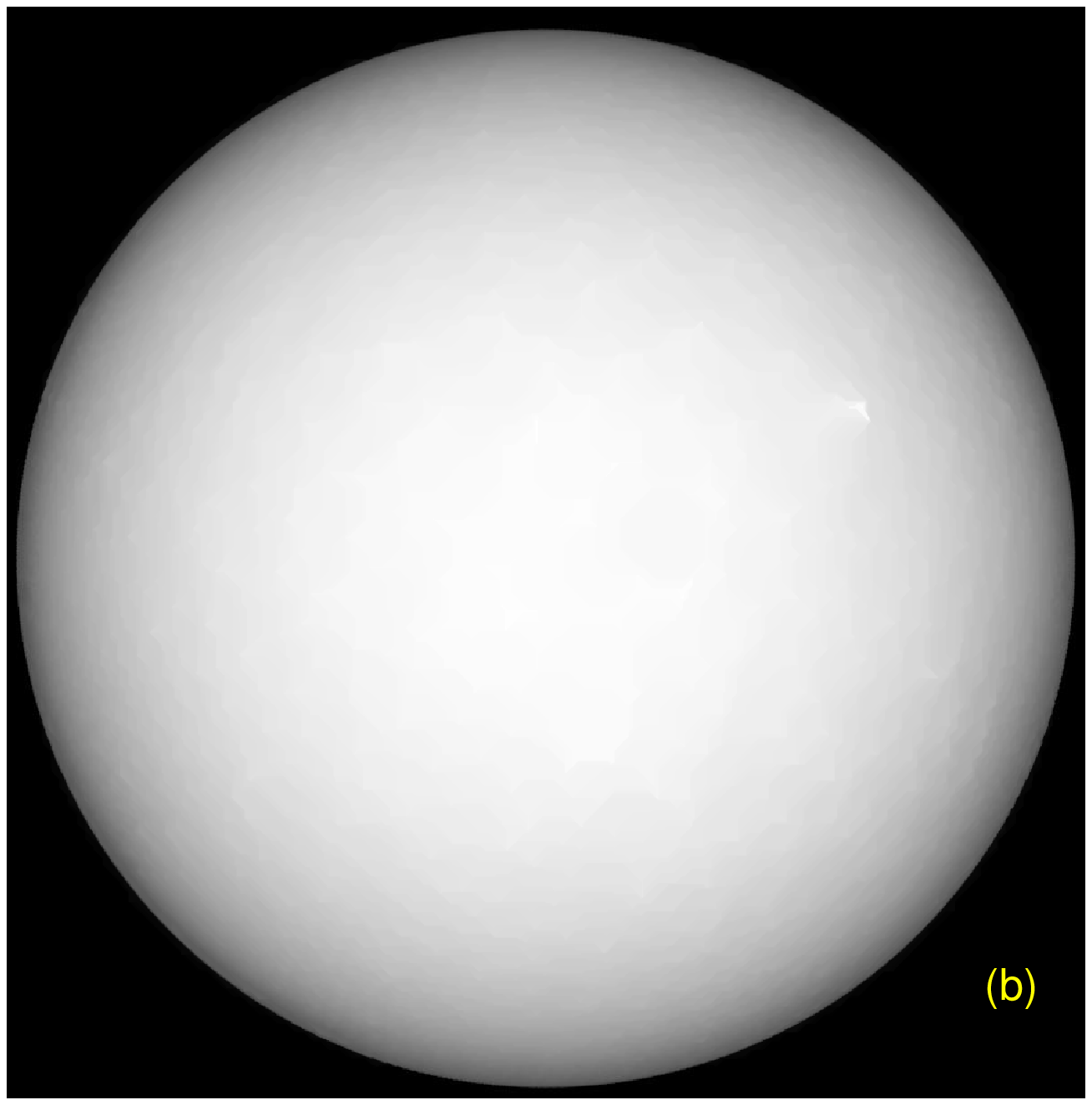}
\includegraphics[width=12pc]{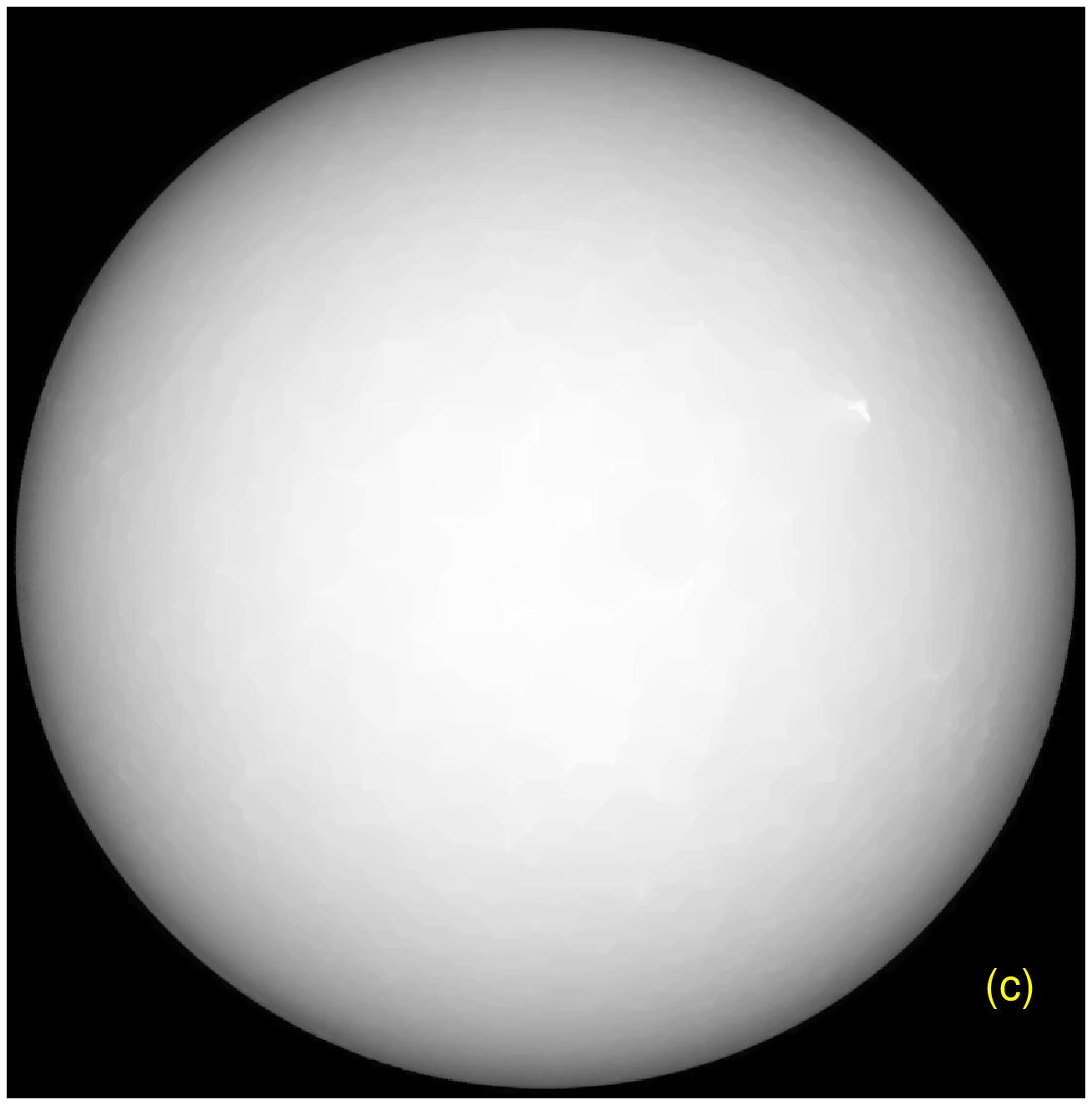}
\includegraphics[width=12pc]{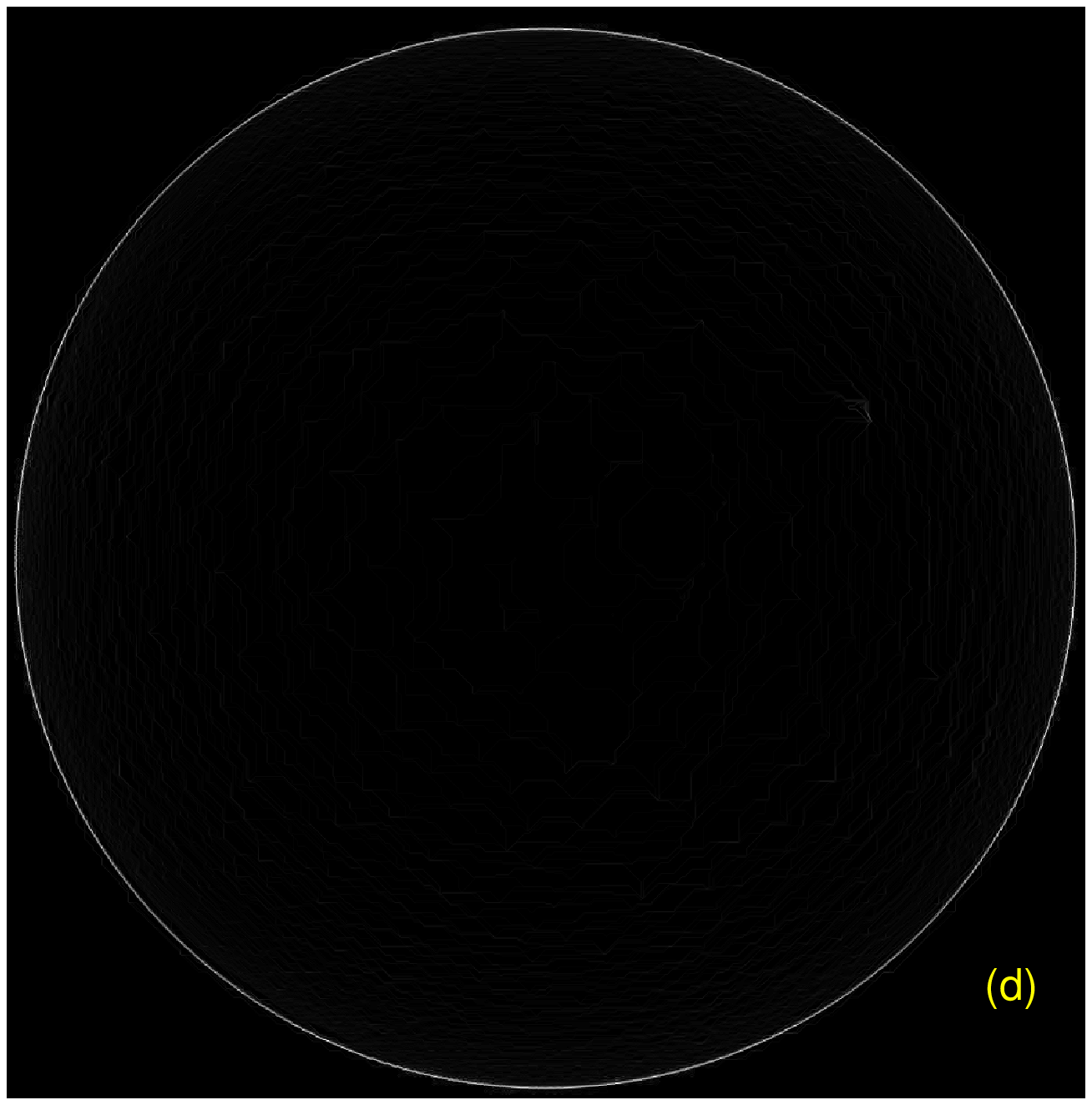}
\includegraphics[width=12pc]{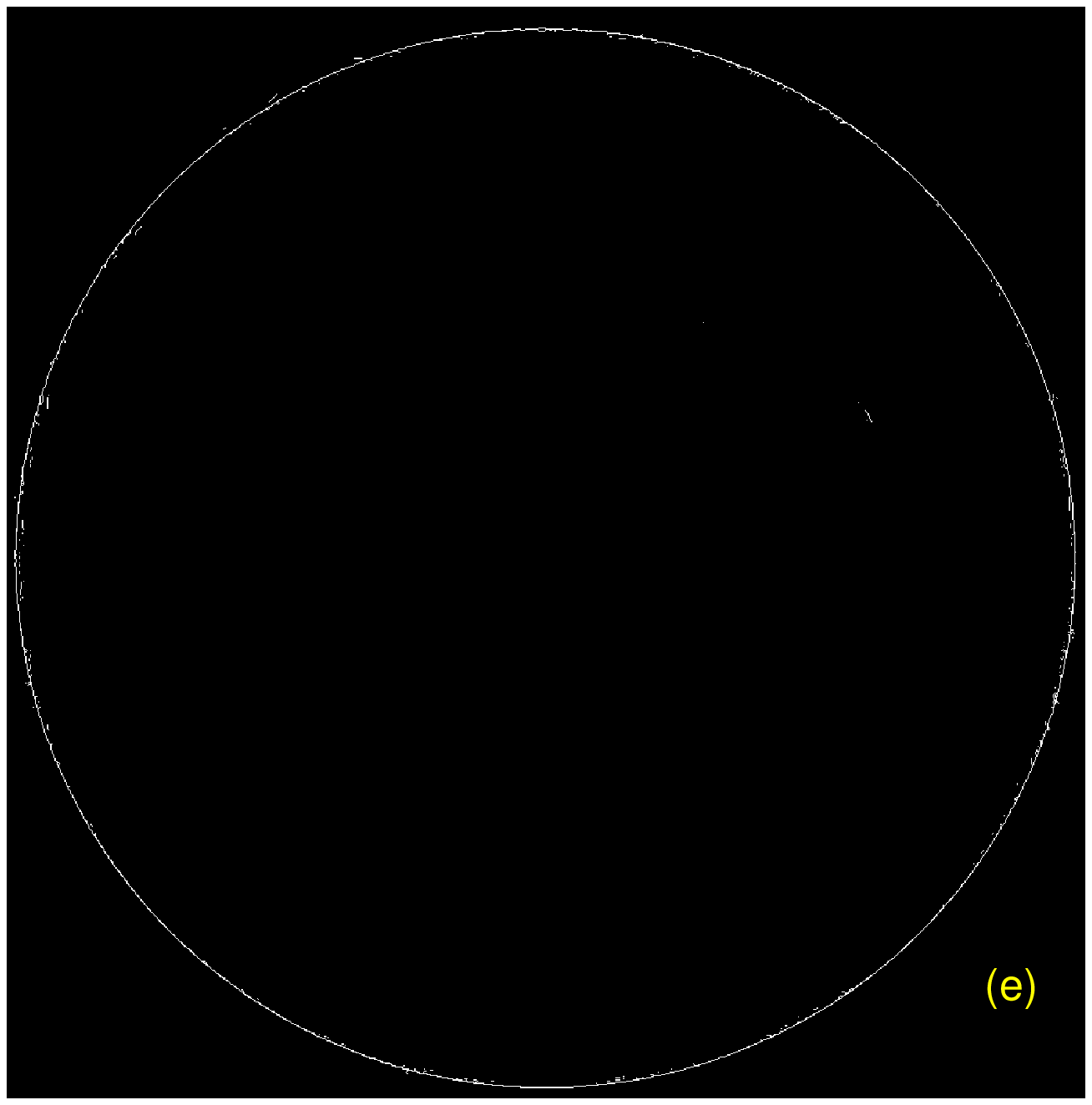}
\includegraphics[width=12pc]{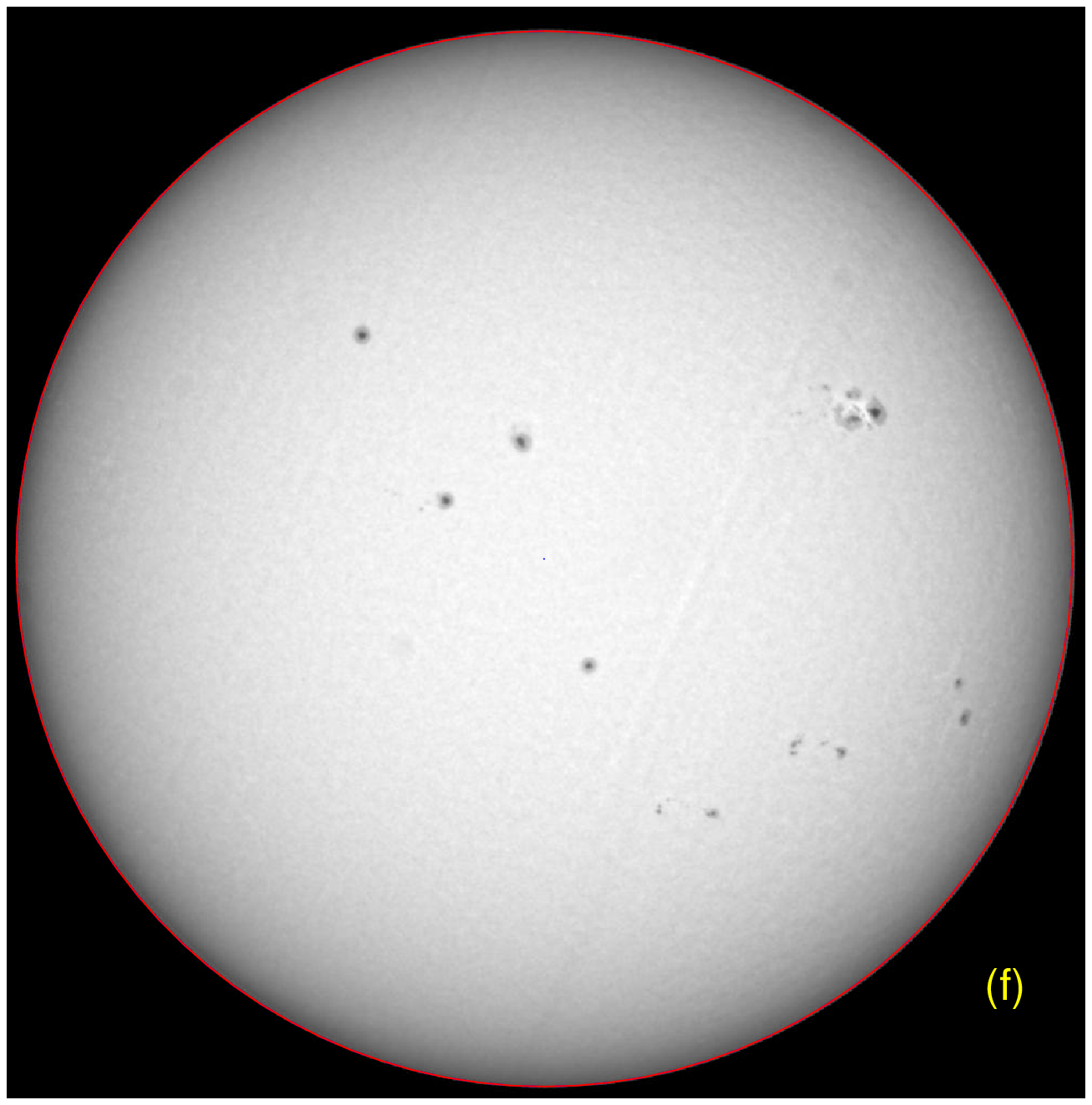}
\caption{A sample of the solar limb extraction in HSOS full-disk photospheric images: (a) the original image; (b) the clean image; (c) the image shrunk of the solar disk, the radius is 1 pixel smaller than that in (b); (d) the solar limb shown in grey image; (e) the solar limb shown in the binary image; (f) the solar limb labelled in red and overlapped on (a).}
 \label{Fig1}
\end{center}
\end{figure*}

To calculate sunspot positions and areas on the solar disk, we must determine the solar centre and radius. This is achieved by extracting the solar limb. A procedure based on a morphological method and Otsu algorithm is designed. The steps are as follows (Figure~\ref{Fig1}):

(1) The first step is to remove sunspots and noises on the solar disk to get a clean solar disk. This is achieved by applying a closing operation (first dilation then erosion) with a structuring element (SE) to the original image in Figure \ref{Fig1}(a). When this SE is larger than sunspots and noises, the closing operation will be able to remove them. The biggest sunspot in hundreds of images is chosen, and its radius calculated to be 30 pixels, so the SE radius is set to be 30. Then a clean solar disk is gotten in Figure \ref{Fig1}(b).

(2) The second step is to shrink the solar disk in Figure \ref{Fig1}(b). To do this, an erosion operation is employed by using an SE. This SE radius is set to be 1 pixel so as to make the radius of the solar disk 1 pixel smaller, then a shrunk solar disk is shown in Figure \ref{Fig1}(c).

(3) To get the solar limb, Figure \ref{Fig1}(c) is subtracted from Figure \ref{Fig1}(b) and \ref{Fig1}(d) is gotten.

(4) The following step is to get the pixel position of the solar limb. Otsu algorithm is applied to segment Figure \ref{Fig1}(d), and a binary image is gotten in Figure \ref{Fig1}(e), on which the white pixels are the position of the solar limb.

(5) In Figure \ref{Fig1}(e), some CCD noises often exist near image border. In order to separate them from the solar limb, a criterion is made, that is as follows: Figure \ref{Fig1}(e) is set to $I$, and its size is $M\times N$, a pixel in the image is set to $I(x,y)$, if $(\frac{N}{2} - y)^2 + (\frac{M}{2} - x)^2 > M^2$, then $I(x,y)=0$. Then CCD noises will be removed from Figure \ref{Fig1}(e).

(6) In the final step, the white pixel positions in Figure \ref{Fig1}(e) are extracted and saved for $X$ and $Y$ arrays, then least square fittingmethod is used to arrays $X$ and $Y$ to create a circle which is considered to be the solar limb. It is labelled with red colour and overlapped on the original image shown in Figure \ref{Fig1}(f).

\subsection{Recognition of sunspots}

\begin{figure*}
\begin{center}
\includegraphics[width=12pc]{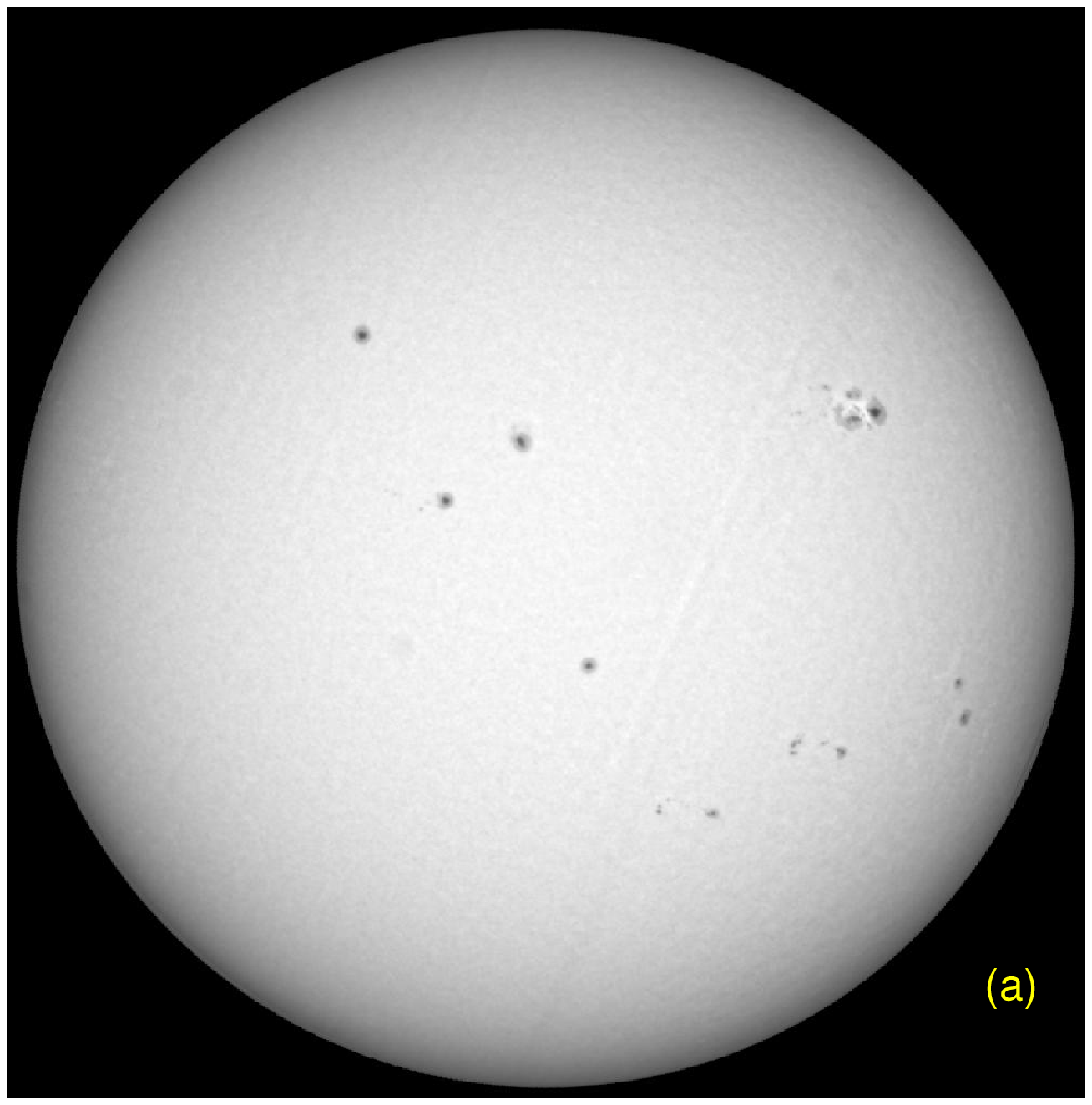}
\includegraphics[width=12pc]{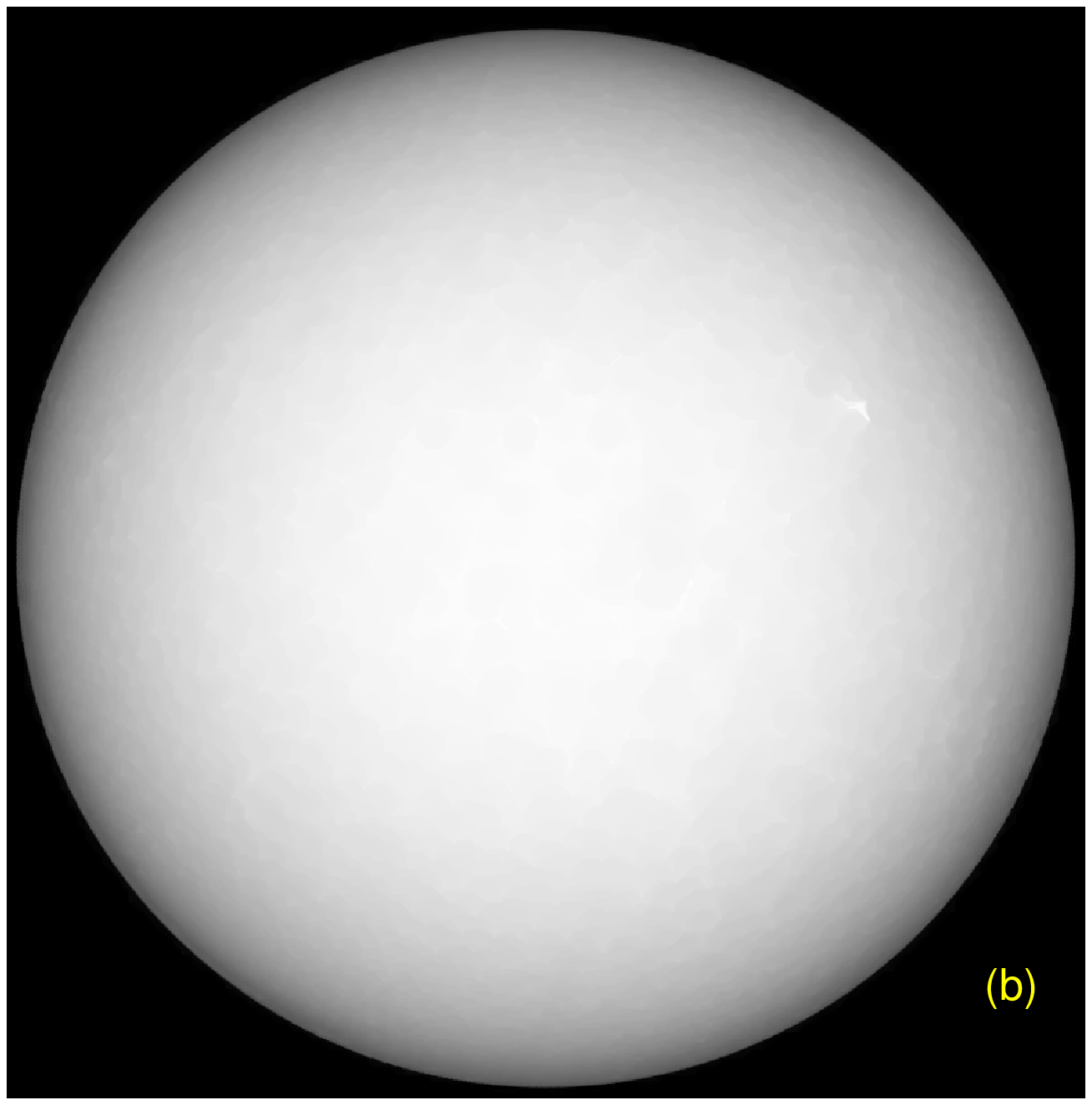}
\includegraphics[width=12pc]{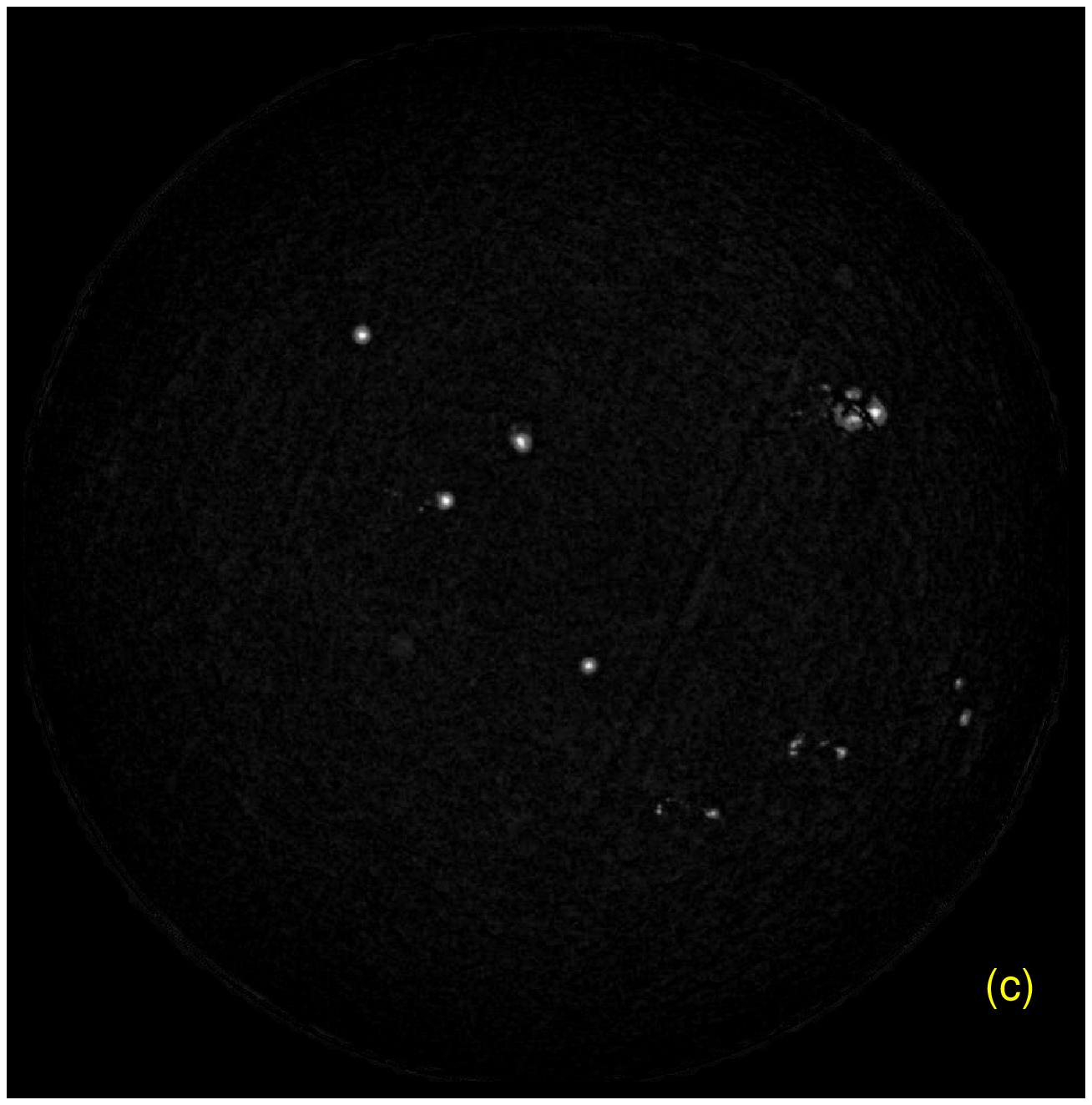}
\includegraphics[width=12pc]{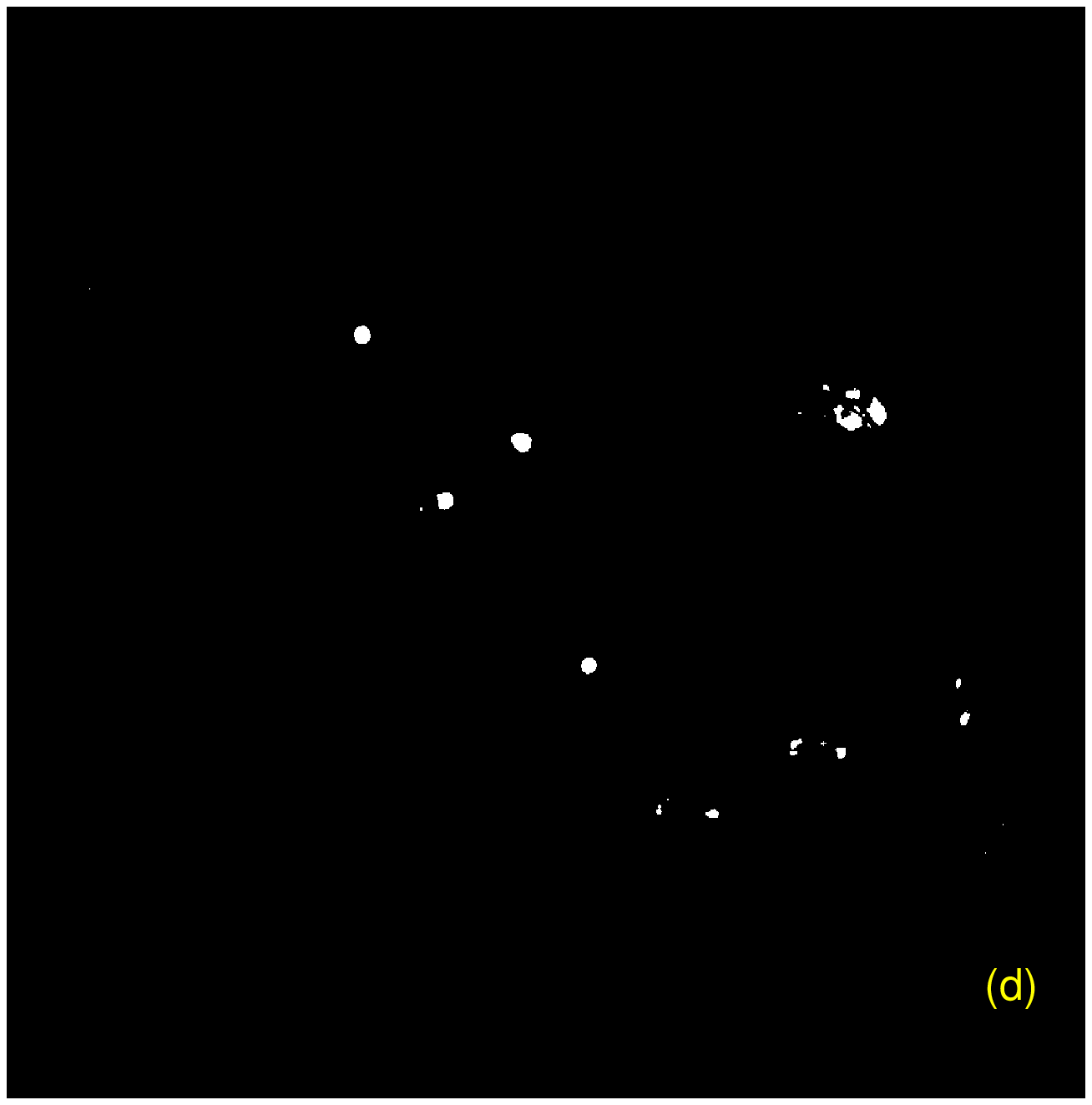}
\includegraphics[width=12pc]{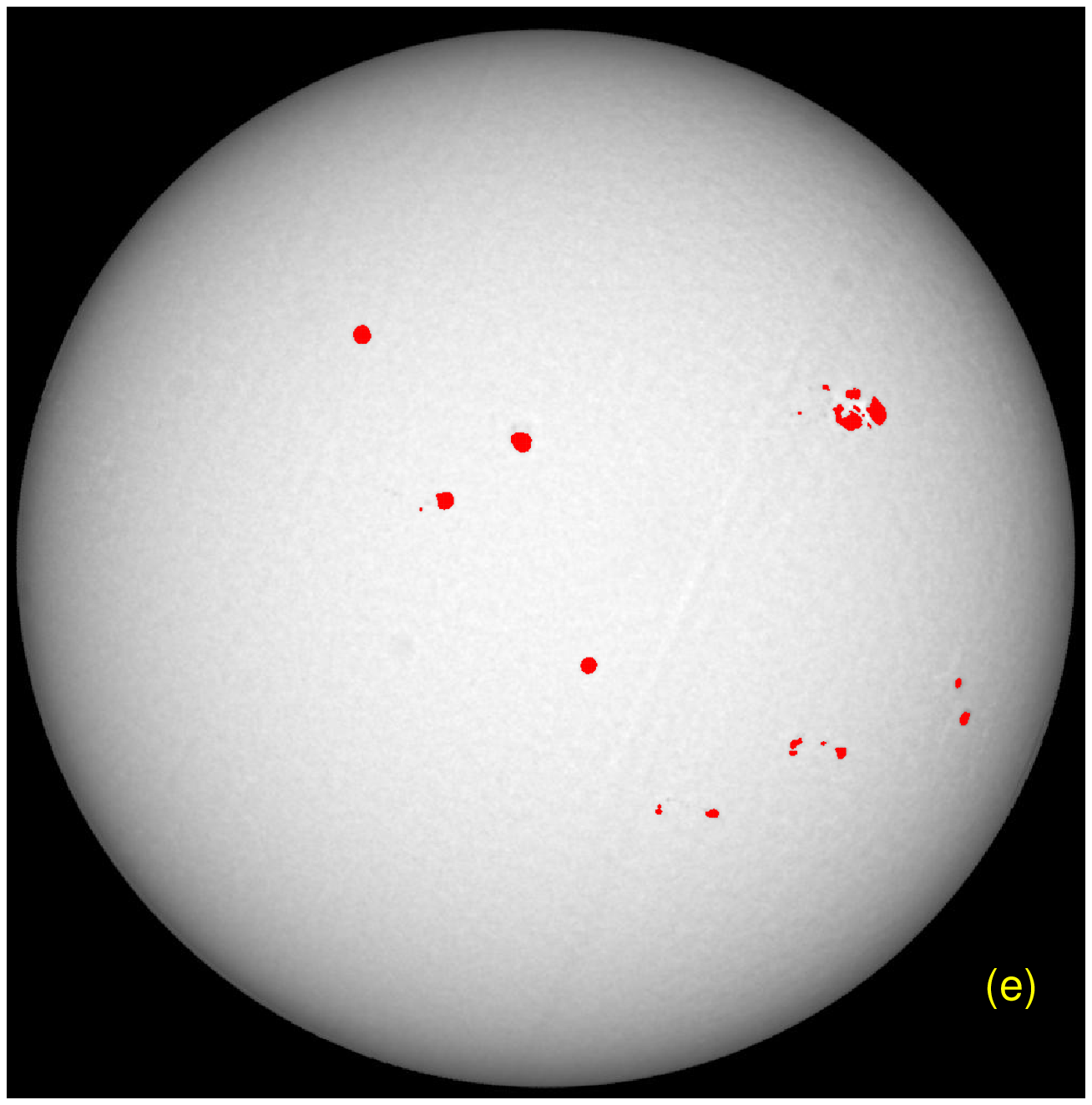}
\caption{The procedure of sunspot recognition in HSOS full-disk photosphere images: (a) the original image; (b) the clean image; (c) the gradient on the image; (d) the binary image showing sunspots candidates; (e) recognised and superimposed sunspots on the original image.}
 \label{Fig2}
\end{center}
\end{figure*}

Once the solar limb is extracted, we can recognise the sunspots in its interior. Due to the limited resolution of data, the sunspot umbra and penumbra could not be separated in HSOS images, so they will be treated as a whole in the processing. The recognition of sunspots is achieved by the following steps (Figure~\ref{Fig2}):

(1) Pre-processing: the method in \citet{WangXF08} is employed to pre-process the original image and Figure~\ref{Fig2}(a) is gotten.

(2) The second step is to get the gradient of the boundaries of sunspots and noises on Figure \ref{Fig2}(a). First, a closing operation with the SE radius of 30 pixels is applied in Figure \ref{Fig2}(a), which is larger than the radiuses of sunspots and noises so as to remove them, the definition of this SE is the same with the method in the first step of Section 4.1, and a clean solar image in Figure \ref{Fig2}(b) is available. Then Figure \ref{Fig2}(a) is subtracted from Figure \ref{Fig2}(b), the gradient is shown in the resulting image of Figure \ref{Fig2}(c).

(3) The third step is to separate the gradient of sunspots from noises in Figure \ref{Fig2}(c). Here, the definition of threshold is the key, it depends on the darkness of Figure \ref{Fig2}(c). By tests and statistics, the suitable value is 20\% of the intensity range of Figure \ref{Fig2}(c). But due to solar limb darkening, the sunspots gradient is lower at the solar limb, we make this threshold smaller to be 15\% on the region of 0.8R of the solar disk (R represents the solar radius). Then sunspots candidate regions are gotten in Figure \ref{Fig2}(d).

(4) The final stage is to acquire sunspots from candidates in Figure \ref{Fig2}(d). We consider the candidates as verified sunspots in which the difference between the max grey value of a pixel and the min one is bigger than 5, and other regions are discarded.

(5) The sunspots are labelled in red and superimposed on the original image. The result is shown in Figure \ref{Fig2}(e).

\subsection{Sunspots recognition in images with instrument noises}

\begin{figure*}
\begin{center}
\includegraphics[width=12pc]{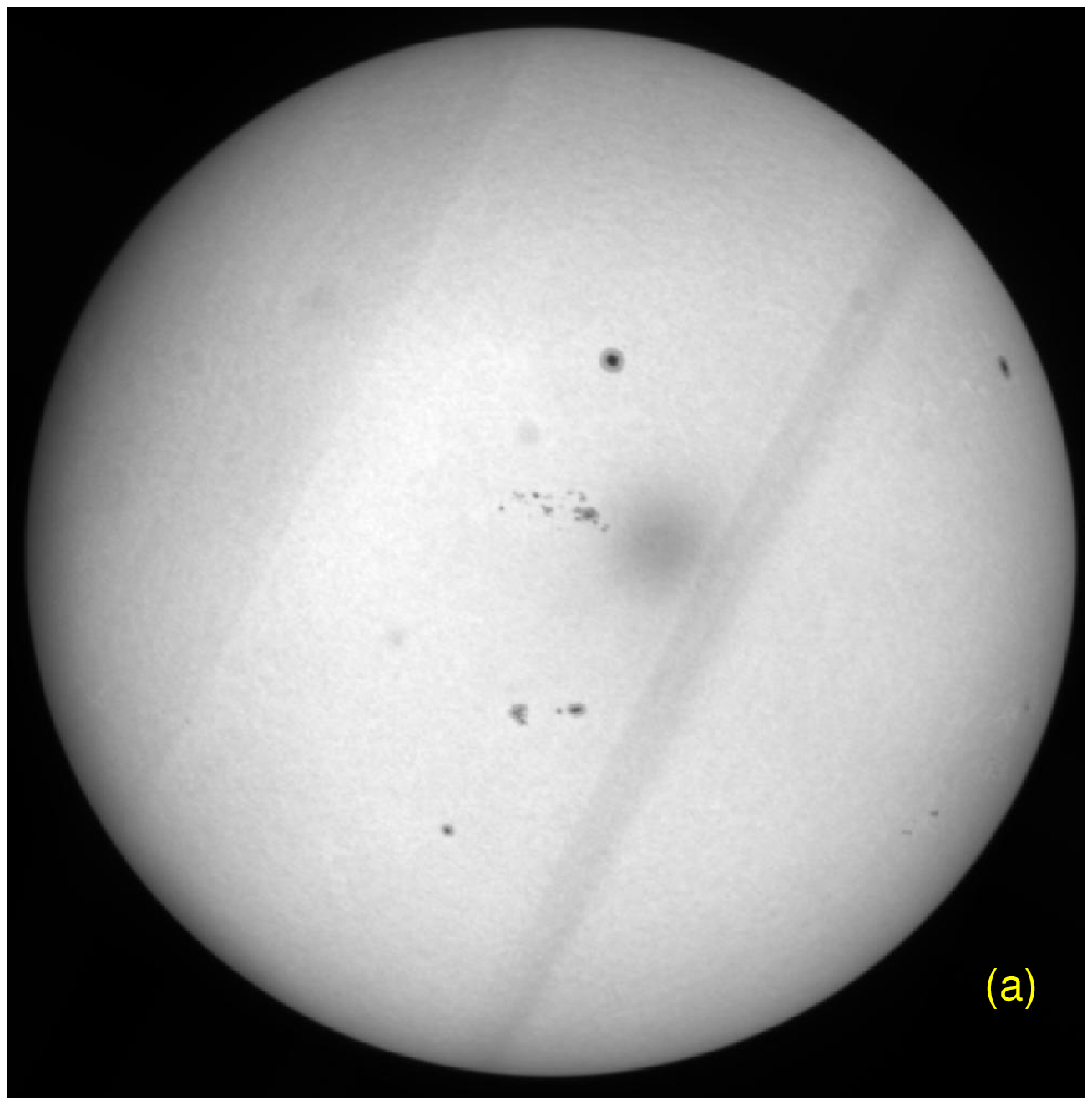}
\includegraphics[width=12pc]{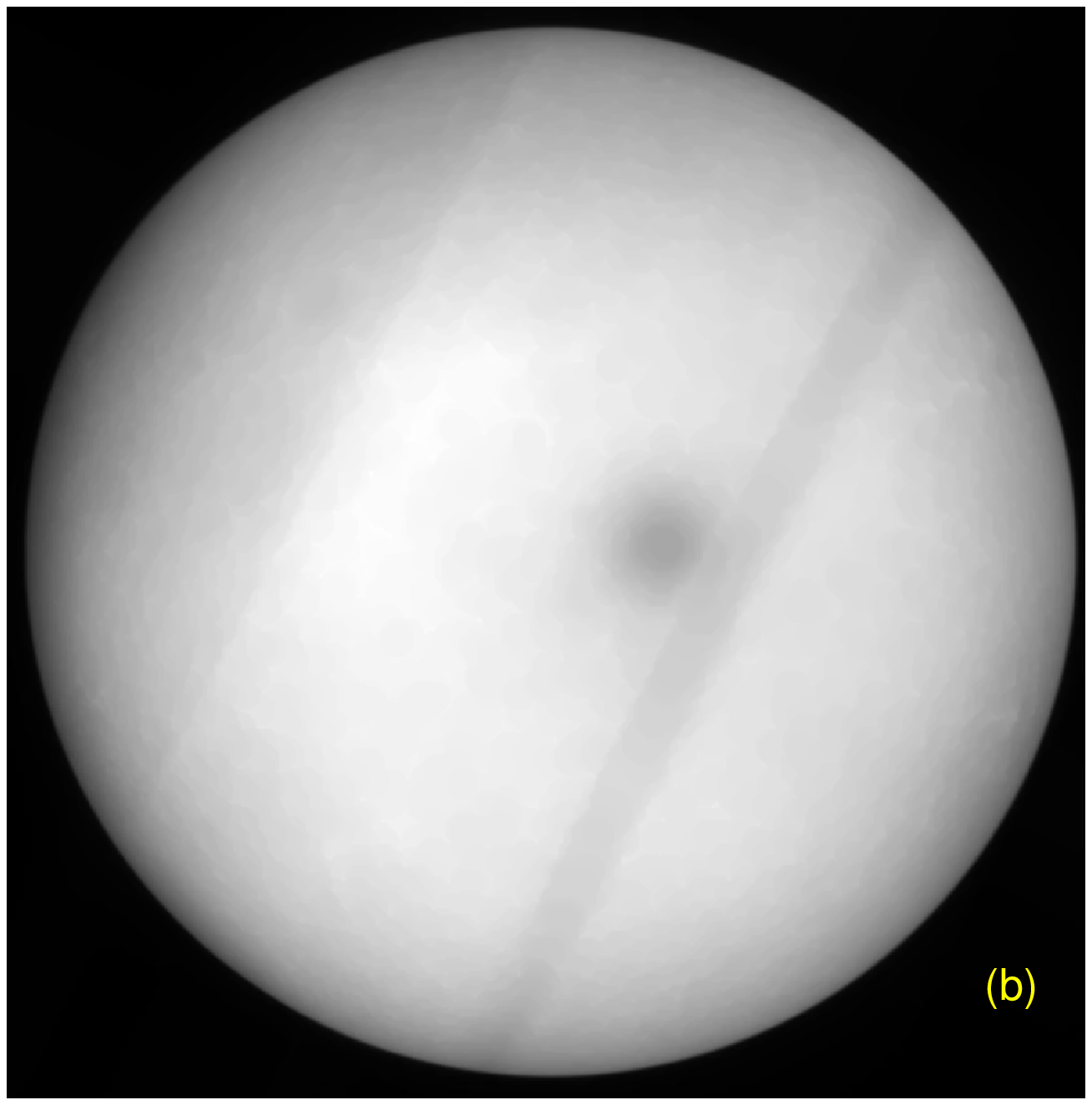}
\includegraphics[width=12pc]{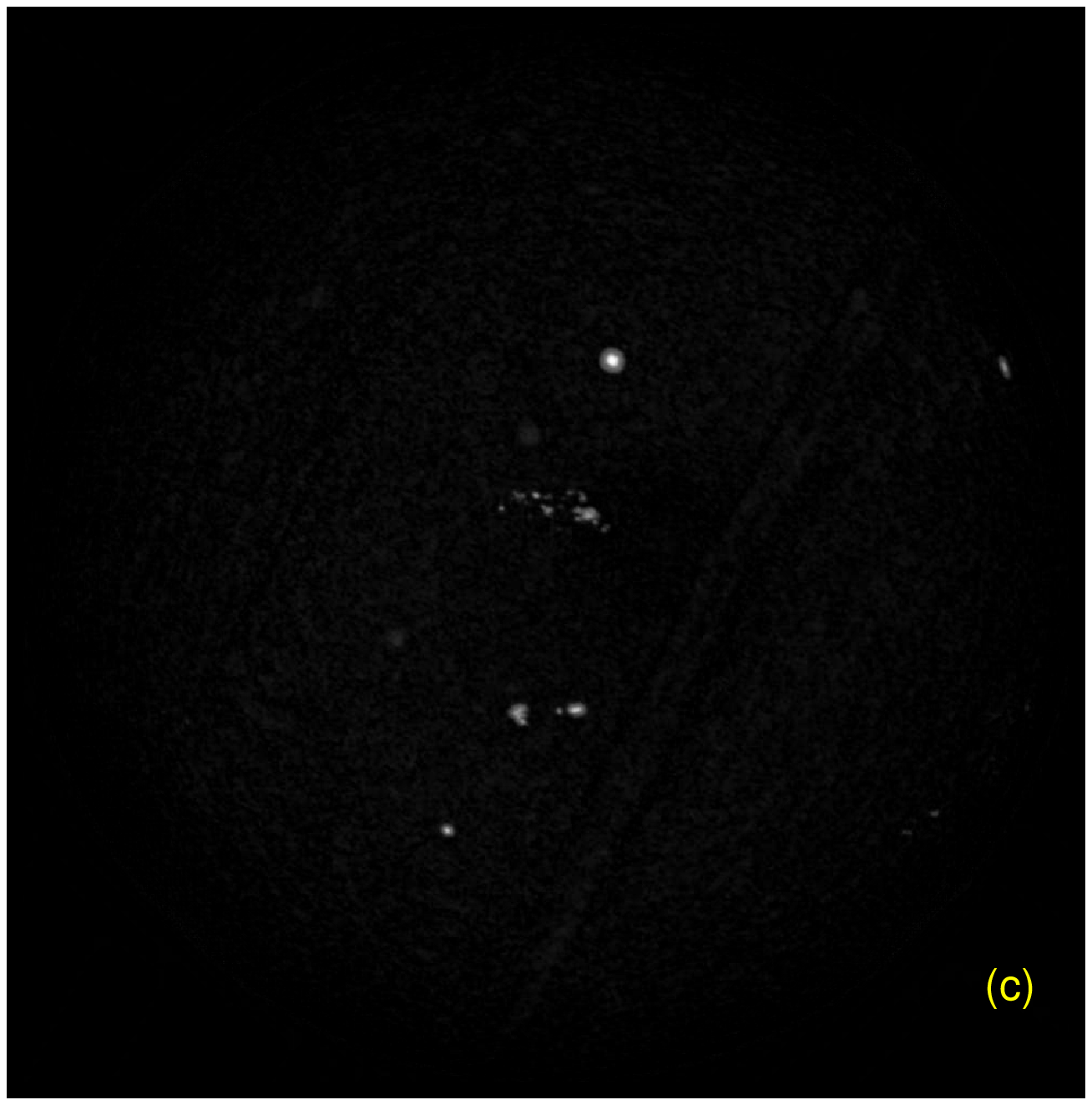}
\includegraphics[width=12pc]{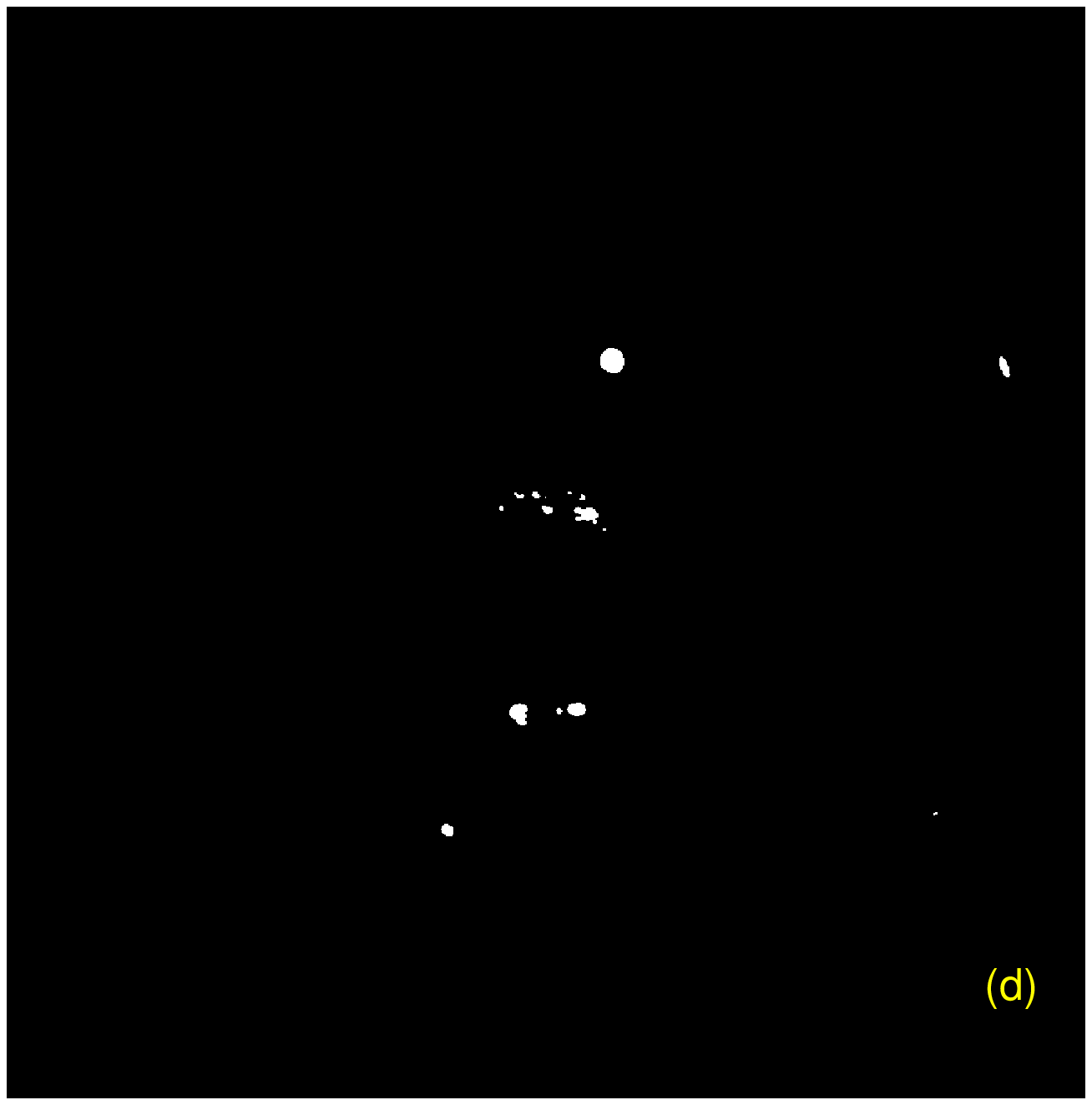}
\includegraphics[width=12pc]{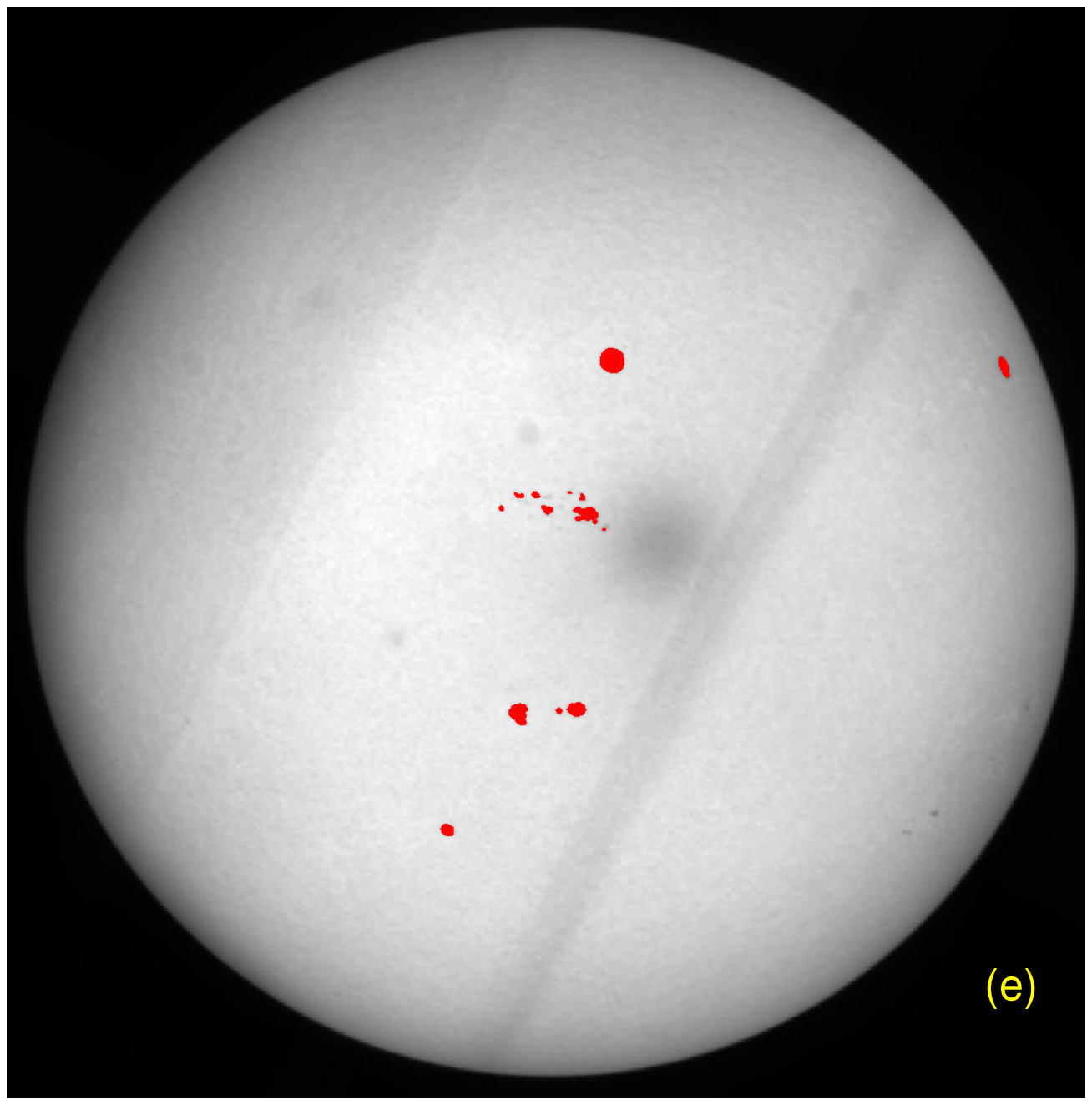}
\caption{(a) The original image disturbed by instrument noises; (b) the clean image without sunspots; (c) the gradient on the image; (d) the binary image showing sunspots candidates; (e) recognised and superimposed sunspots on the original image.}
 \label{Fig3}
\end{center}
\end{figure*}

As mentioned in the first paragraph, the instrument noises which appear in HSOS images make the previous methods not work well. So this procedure is used to test the effect of the recognisation of the sunspots. The result is shown in Figure~\ref{Fig3}, in which we can see the procedure performs well for these images.

\section{VERIFICATION AND DISCUSSION}

Two methods are adopted to verify the accuracy of the procedure. The first is to detect sunspots by our automatic procedure and manual method separately based on the existed data set, and make a comparison between them. The second is to compare two different data sets taken in the same period, we calculate the correlations of the sunspot areas taken from them by the automatic procedure. More details and results are given below.

\begin{table*}
\caption{The accuracy of sunspot recognition by the automatic procedure in comparison with manual one.}
\begin{center}
\begin{tabular*}{\textwidth}{@{}c\x c\x c\x c\x c\x c\x c\x c\x c\x c\x c\x c@{}}
\hline \hline
 Date   & number of spots   &  number of spots  &  FRR  & FAR  \\
        &  {(manual method)}   &   {(the automatic method)}   \\
\hline
 2011-11-2 &10 &9  &1 &0 \\
 2011-11-6 &12 &12 &0 &0   \\
 2011-11-7 &17 &17 &0 &0  \\
 2011-11-9 &15 &15 &0 &0 \\
 2011-11-10 &15 &15 &0 &0 \\
 2011-11-11 &18 &18 &0 &0 \\
 2011-11-12 &15 &14 &1 &0 \\
 2011-11-14 &15 &15 &0 &0 \\
 2011-11-18 &11 &8 &3 &0 \\
 2011-11-19 &9 &8 &1 &0 \\
 2011-11-20 &13 &12 &1 &0 \\
 2011-11-21 &9 &7 &2 &0 \\
 2011-11-22 &13 &12 &1 &0 \\
 2011-11-23 &11 &10 &1 &0 \\
 2011-11-25 &14 &13 &1 &0 \\
 2011-11-30 &17 &17 &0 &0 \\
 2011-12-1 &12 &10 &2 &0 \\
 2011-12-3 &14 &12 &2 &0 \\
 2011-12-4 &22 &24 &0 &2 \\
 2011-12-7 &12 &12 &0 &0 \\
 2011-12-8 &11 &11 &0 &0 \\
 2011-12-9 &7 &7 &0  &0 \\
 2011-12-10 &8 &8 &0 &0 \\
 2011-12-12 &3 &4 &0 &1 \\
 2011-12-14 &5 &4 &1 &0 \\
 2011-12-15 &4 &4 &0 &0 \\
 2011-12-16 &5 &5 &0 &0 \\
 2011-12-19 &11 &11 &0 &0 \\
 2011-12-20 &6 &6 &0 &0  \\
 2011-12-21 &10 &9 &1 &0 \\
 2011-12-22 &10 &11 &0 &1 \\
 2011-12-23 &9 &9 &0 &0 \\
 2011-12-24 &6 &5 &1 &0 \\
 2011-12-25 &8 &8 &0 &0 \\
 2011-12-26 &10 &10 &0 &0 \\
 2011-12-28 &9 &8 &1 &0 \\
 2011-12-31 &7 &8 &0 &1 \\
 SUM &403 & 388 &20   &5\\
\hline \hline
\end{tabular*}\label{tab1}
\end{center}
\tabnote{FRR: the number of sunspots detected by the manual but not by the automatic.}
\tabnote{FAR: the number of sunspots detected by the automatic but not by the manual.}
\end{table*}

\begin{table*}
\caption{Diameter of sunspots not recognised by our automatic method.}
\begin{center}
\begin{tabular}{@{}cc@{}}
\hline\hline
Unrecognized sunspot number  & Diameter  \\
\hline%
 1 &1.99   \\
 2 &1.24   \\
 3 &1.95   \\
 4 &0.98   \\
 5 &1.05   \\
 6 &1.67   \\
 7 &1.00   \\
 8 &0.87   \\
 9 &1.77   \\
 10 &1.39   \\
 11 &1.39   \\
 12 &1.61   \\
 13 &1.45   \\
 14 &1.54   \\
 15 &1.15   \\
 16 &1.01   \\
 17 &1.05   \\
 18 &0.98   \\
 19 &1.01   \\
 20 &1.48   \\
\hline\hline
\end{tabular}
\end{center}
\label{tab2}
\end{table*}

\subsection{Accuracy of automatic procedure compared with manual}

Sunspot number is a basic index to reflect the level of solar activities, and is a suitable vehicle to assess the accuracy of our procedure \citep{Hoyt98}. In our case, the data set is from HSOS photospheric images from November 2011 to December. The automatic and manual methods are adopted separately to calculate the sunspot numbers automatically and count them manually. Two results are listed in Table~\ref{tab1}. The first column of the table shows the date of taking the images, the second column the number of sunspots counted by the manual method in an image, the third column the number of sunspots done by our automatic procedure, the fourth column is false rejection rate FRR (the number of sunspots detected by the manual but not by the automatic), and the last column is false acceptance rate FAR (the number of sunspots detected by the automatic but not by the manual).

In the process of manual recognition, small sunspots are easily missed due to limited seeing condition. To avoid this, the images of SDO/HMI are used as reference to ensure small sunspots to be detected.

In the last row of the table, we sum the total sunspot numbers recognised by manual and automatic method respectively, and the total FRR and FAR.

Then we define the recognition rate as follows:
\begin{align*}
\frac{\text{\footnotesize{sum of the automatic method}} - \text{\footnotesize{sum of the FAR}}}{\text{\footnotesize{sum of the manual method}}} &= \frac{388 - 5}{403}  \\
&= 95\%,
\end{align*}
and the error rate as follows:
\[
\frac{\text{\footnotesize{sum of the FAR}}}{\text{\footnotesize{sum of the manual method}}} = \frac{5}{403} = 1.2\%
\]

Analysing the sunspots which the automatic method detected but the manual did not, it may be caused by tiny dust or noises, such as instrument noise. The reasons why the sunspots can be detected by the manual but not by the automatic are mainly the following:

(1) For some adhesive sunspots, the manual method will deal with them as separated ones, but the automatic method regards them as one; (2) in our recognition procedure to remove noises, the criterion we set is a region whose difference between maximum grey and minimum one is larger than 5 of being a sunspot. This works well in most cases, except a few cases in which some small sunspots are still missed.

We calculate the diameters of the sunspots which are missed by the automatic method, the diameter calculation formula is
\[
d = \frac{\sqrt{s/\pi}}{R} \times 180\,^{\circ}
\]
where $s$ is the sunspot area, $R$ is the solar disk radius acquired by the automatic solar limb detection program, and $d$ the sunspot diameter in units of degree.

The result is shown in Table~\ref{tab2}. From the table, we can see their diameters are all less than 2$\,^{\circ}$, which means they are weak sunspots on the solar disk which have little impact on the level of solar activities.

\subsection{Verification with USAF/NOAA}

Sunspot area is an important indicator of the solar activity level \citep{Carbonell92}, which is associated with the solar cycle and significant in space environment monitoring. To further verify the automatic method, we make use of the sunspot area data in this paper.

\begin{figure*}
\begin{center}
\includegraphics[width=\columnwidth]{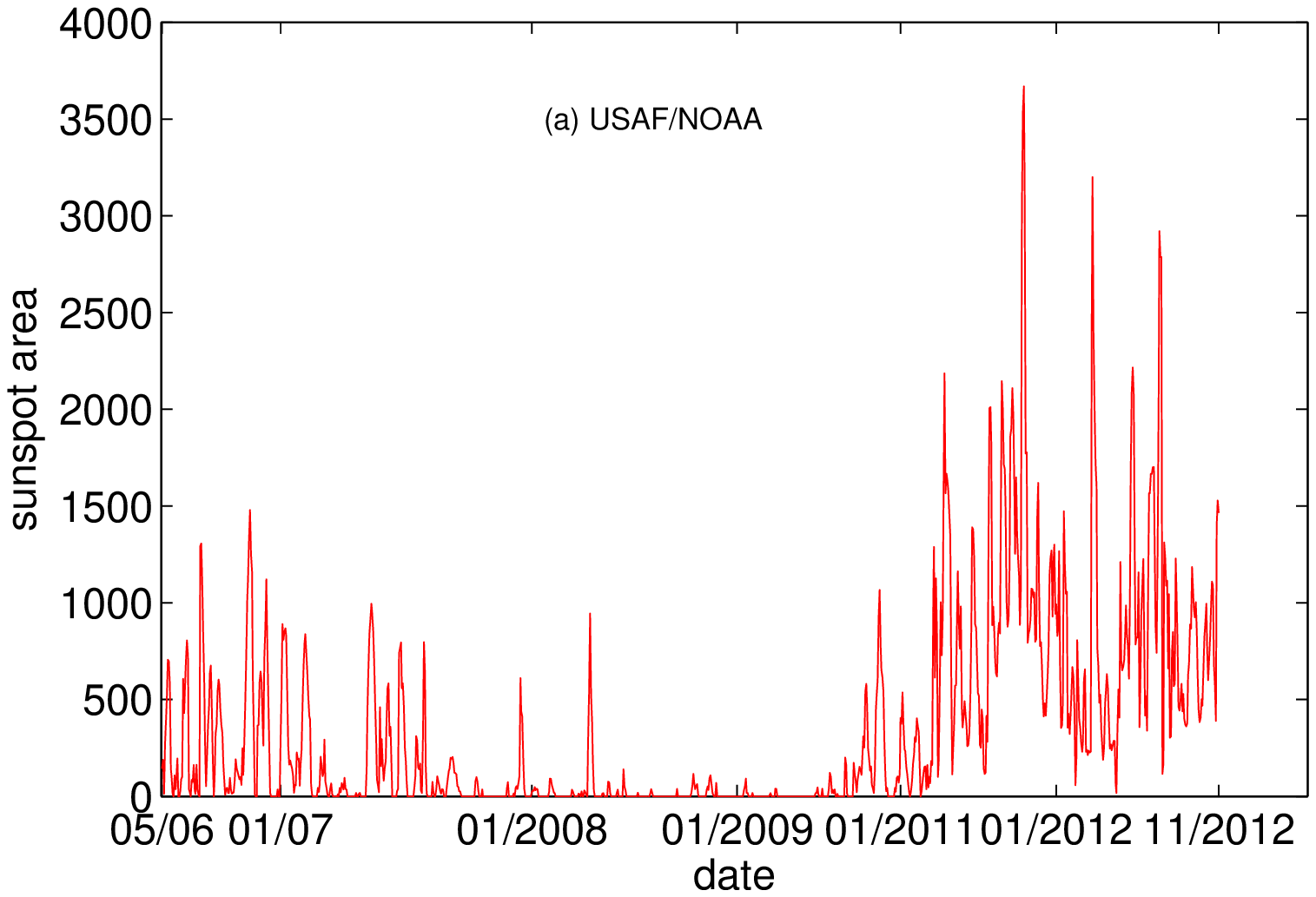}
\includegraphics[width=\columnwidth]{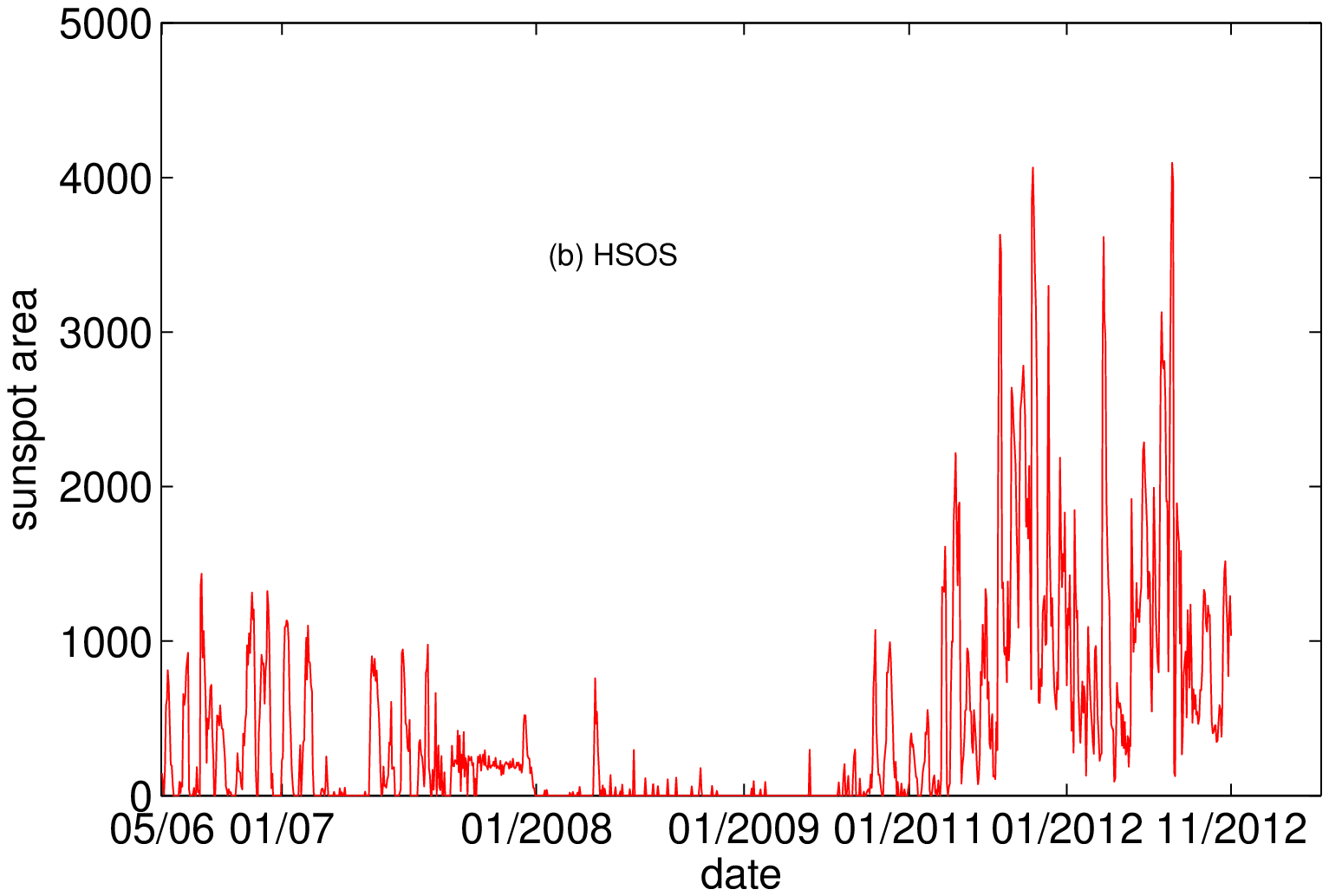}
\caption{(a) Sunspot areas provided by USAF/NOAA in 2006--2012; (b) sunspot areas extracted from HSOS by the automatic method in 2006--2012.}
\label{Fig4}
\end{center}
\end{figure*}

Figure~\ref{Fig4} shows the comparison of the daily sunspot areas extracted by the automatic method from HSOS images during the period of 2006¨C-2012 with those available as TXT files at the USAF/NOAA. The horizontal axis represents the date of taking images, the vertical the sunspot area, in units of millionths of a solar hemisphere ($\mu$Hem). Due to the maintenance of the instrument from August 2009 to November 2010, HSOS data are blank.

\begin{figure*}
\begin{center}
\includegraphics[width=\columnwidth]{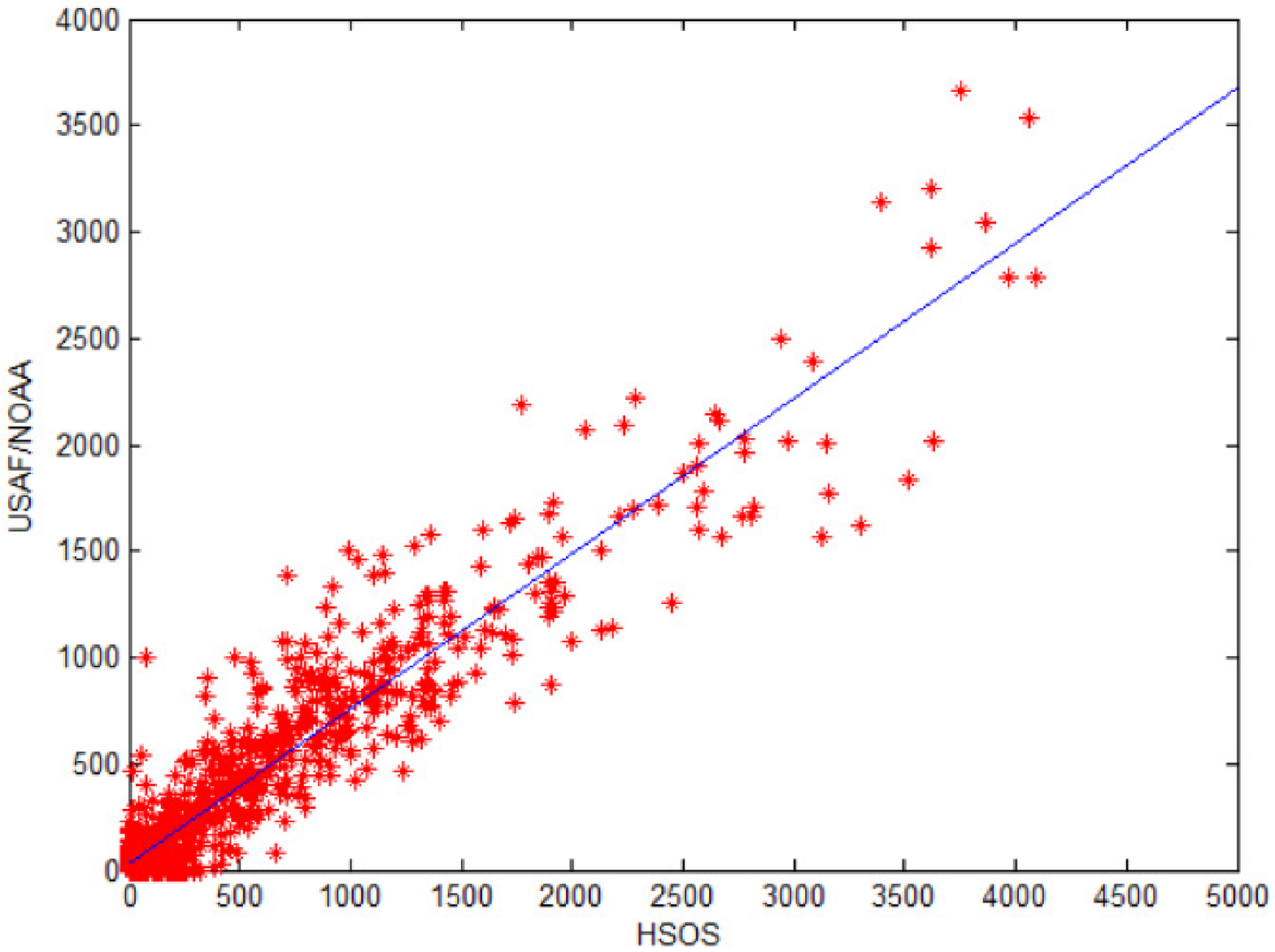}
\caption{Correlation between USAF/NOAA and HSOS sunspots areas.}
\label{Fig5}
\end{center}
\end{figure*}

From Figure~\ref{Fig4}, we can see that USAF/NOAA and HSOS have the same tendency, with the correlation coefficient of 95\% shown in Figure~\ref{Fig5}. This is a high accuracy of recognition which could prove the procedure works well.

To analyse the difference between Figure \ref{Fig4}(a) and \ref{Fig4}(b),we find the discrepancy is caused by a few factors: first, some instrument noises indistinguishable from smaller sunspots lead to false identification; second, atmospheric interference to HSOS images makes the sunspots border more indistinct; finally, although we choose and compare the images of HSOS and USAF/NOAA in the same days, their collection time may be different that the sunspot morphology may somewhat change.

\subsection{Enlarge the recognised sunspots}

\begin{figure*}
\begin{center}
\includegraphics[width=10pc]{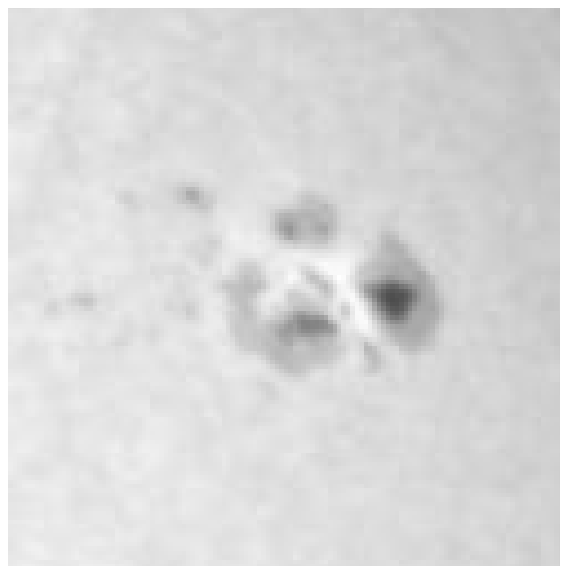}
\includegraphics[width=10pc]{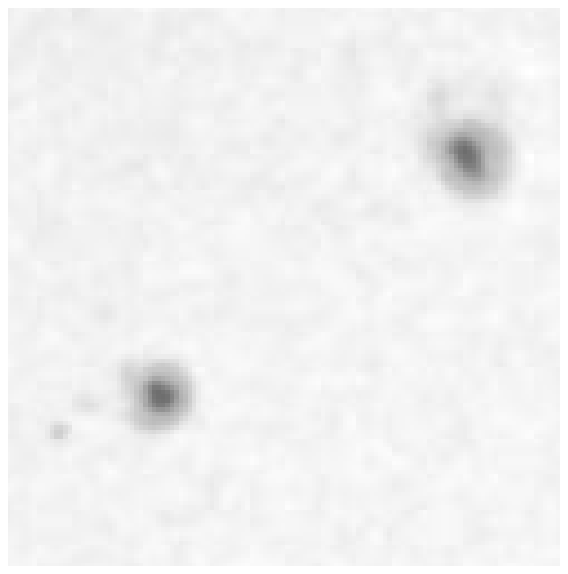}
\includegraphics[width=10pc]{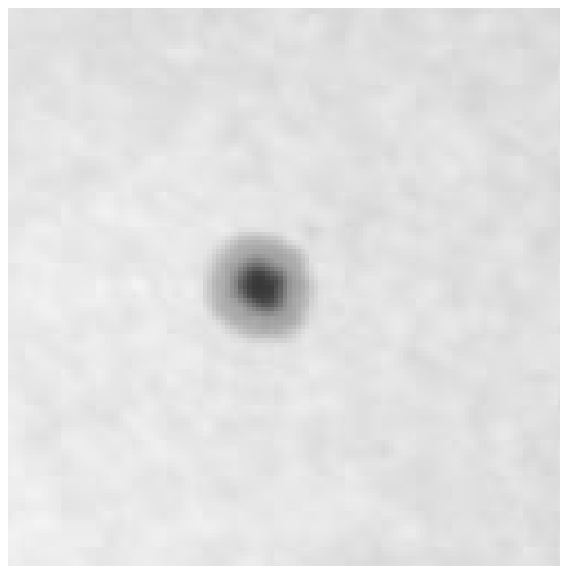}
\includegraphics[width=10pc]{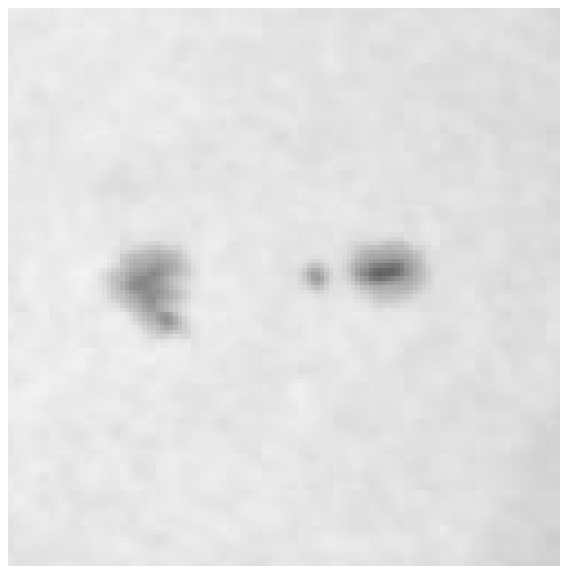}
\includegraphics[width=10pc]{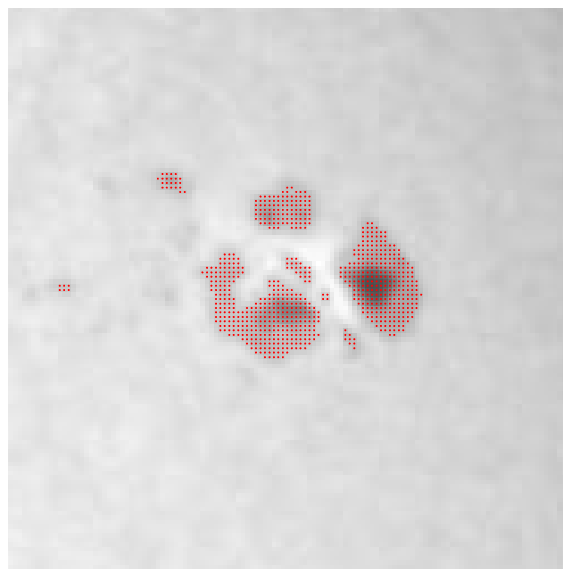}
\includegraphics[width=10pc]{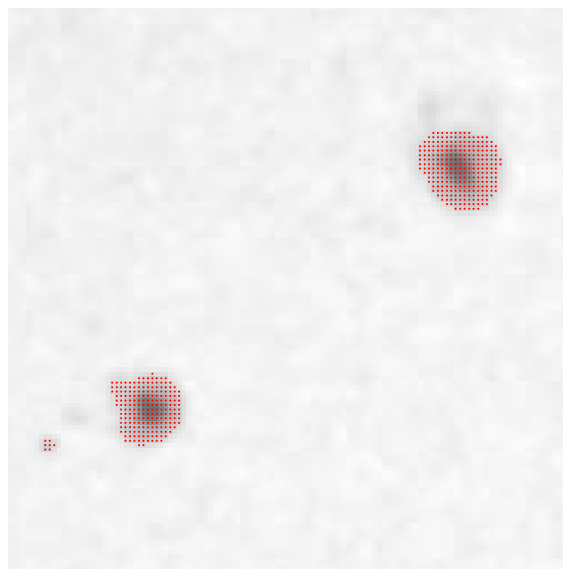}
\includegraphics[width=10pc]{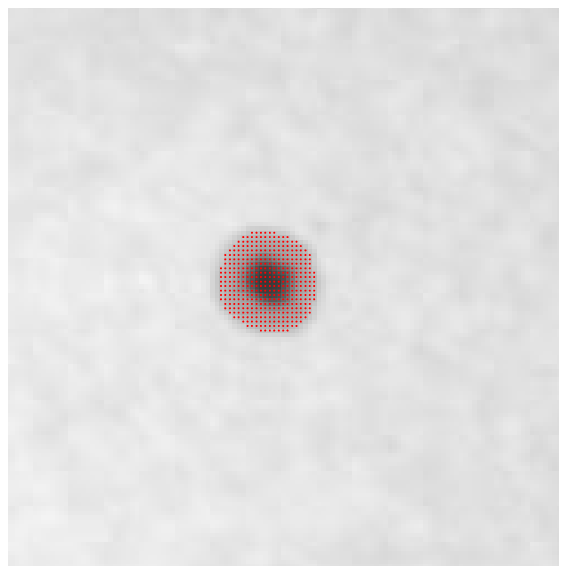}
\includegraphics[width=10pc]{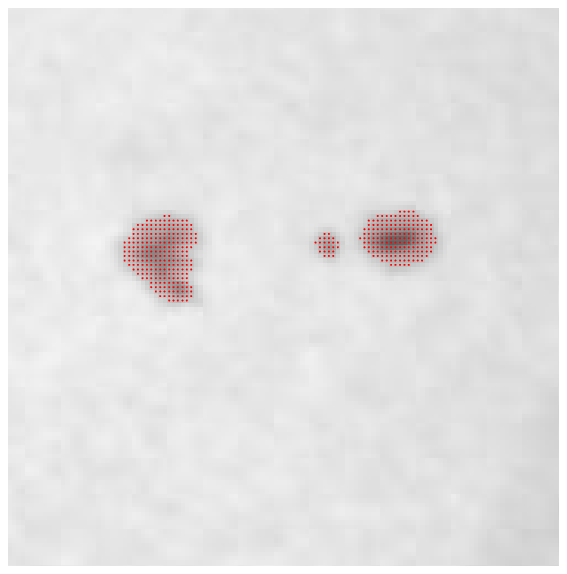}
\caption{Enlarge the recognised sunspots: the first line shows the original image of sunspots; the second line shows the zoomed detected areas.}
 \label{Fig6}
\end{center}
\end{figure*}

In order to show the detection result, some sunspots are selected randomly and recognised. The detected areas are zoomed and compared with the original images, they are shown in Figure~\ref{Fig6}. From the figure ,it seems that the sunspots are recognised accurately.

\section{CONCLUSION}

A procedure for recognition of the sunspots in full-disk photospheric solar images of HSOS is introduced. It adopts Gaussian algorithm to smooth the images at first. Then the sunspots are recognised through the morphological Bot-hat operation with a local threshold. Wrong selection of sunspots is eliminated by a criterion of limiting sunspot properties. Besides, the morphological operations and Otsu algorithm are used to extract the solar limb, which is helpful to calculate sunspot areas. Compared with the manual method, the recognition rate is 95\%, the error rate is 1.2\%. The correlation between USAF/NOAA and HSOS sunspots areas is in a good agreement (95\%). The advantage of this procedure is that it is appropriate to detect sunspots for lower resolution images, particularly the images associated with instrument noises.

In the next step, we will focus on improving the accuracy of recognition. A new sunspot property database will be expected and released on web site, which will be helpful for a study of the solar activity.

\begin{acknowledgements}
One of the authors C. Zhao is grateful to Prof. X. J. Mao for giving beneficial advices for the manuscript. Thiswork is supported by National Natural Science Foundation of China (11427901, 11403058, 11473039, 11178005, U1331113, U1531247, 11203036, U1331104, 11427803, 11303052, 11373040, 11573037), Ministry of Science and Technology of the People¡¯s Republic of China (2012FY120500, 2014FY120300), and the Young Researcher Grant of National Astronomical Observatories, Chinese Academy of Sciences.
\end{acknowledgements}


\nocite*{}
\bibliographystyle{pasa-mnras}
\bibliography{zhaoc}

\end{document}